\documentclass[journal]{IEEEtran}
\usepackage{times,amsfonts,bbm,dsfont,amsmath,color,amssymb,graphicx,epsfig,multirow,float,algorithm,algorithmic,bm,cite,setspace}
\usepackage[normalem]{ulem}
\usepackage{makecell}
\usepackage{stfloats}
\usepackage{arydshln}
\usepackage{psfrag}
\usepackage{epstopdf}
\usepackage{stfloats}
\usepackage{color,cases}
\usepackage{subfigure}

\begin{document}

\title{20 Years of Evolution from Cognitive to Intelligent Communications}
\author{
\IEEEauthorblockN{Zhijin Qin, Xiangwei Zhou, Lin Zhang, Yue Gao, Ying-Chang Liang, and Geoffrey Ye Li }

\thanks{This work was supported in part by the National Science Foundation under Grant No. 1560437.}
\thanks{Zhijin Qin and Yue Gao are with Queen Mary University of London, London E1 4NS, UK, (e-mail: z.qin@qmul.ac.uk, yue.gao@qmul.ac.uk).}
\thanks{Xiangwei Zhou is with Louisiana State University, Baton Rouge, LA, USA, 70803, (email: xwzhou@lsu.edu).}
\thanks{Lin Zhang and Ying-Chang Liang are with University of University of Electronic Science and Technology of China, Chengdu 611731, China. (e-mail: linzhang1913@gmail.com, ycliang@uestc.edu.cn).)}
\thanks{Geoffrey Ye Li is with Georgia Institute of Technology, Atlanta, GA, USA, 30332-0250, (e-mail: liye@ece.gatech.edu).}
}
\maketitle

\begin{abstract}
It has been 20 years since the concept of \textit{cognitive radio} (CR) was proposed, which is an efficient approach to provide more access opportunities to connect  massive wireless devices.  To improve the spectrum efficiency, CR enables unlicensed usage of licensed spectrum resources. It has been regarded as the key enabler for intelligent communications. In this article, we will provide an overview on the intelligent communication in the past two decades to illustrate the revolution of its capability from cognition to \textit{artificial intelligence} (AI). Particularly, this article starts from a comprehensive review of typical spectrum sensing and sharing, followed by the recent achievements on the AI-enabled intelligent radio. Moreover, research challenges in the future intelligent communications will be discussed to show a path to the real deployment of intelligent radio. After witnessing the glorious developments of CR in the past 20 years, we try to provide readers a clear picture on how  intelligent radio  could be further developed to smartly utilize the limited spectrum resources as well as to optimally configure wireless devices in the future communication systems.


\end{abstract}

\section{Introduction and challenges}

From the emergence of the \textit{first generation} (1G) of cellular communications in 1979 to the deployment of the \textit{fifth generation} (5G) in 2019, it takes around ten years for the evolution of each generation \cite{david20186g,patzold20195g,agiwal2016next}. The first two generations of wireless communication systems mainly aim to provide reliable  voice services over a wide coverage area, which consumes an acceptable amount of spectrum resource. With the rapid increase in the demands for high data-rate services, the  spectrum resource becomes the bottleneck that constrains the development of wireless communications. To deal with the issue, the engineers and researchers from both industry and academia started to study the intelligent communications after the \textit{second generation} (2G) of cellular communications. In 2000, the concept of  intelligent communications, i.e., \textit{cognitive radio} (CR), was proposed by Mitola in \cite{Mitola:1999}. CR enables the radio devices to learn the radio environment and adapt their configurations to enhance the spectrum utilization.

\begin{figure}[!ht]
\centering
\includegraphics[width=3.0in]{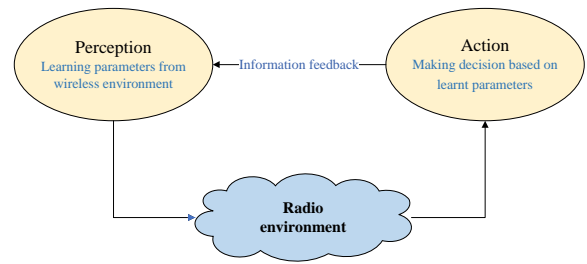}
\caption{Perception-action cycle in intelligent communications.}
\label{Intelligent_decision_making_process}
\end{figure}

Fig. \ref{Intelligent_decision_making_process} shows the famous perception-action cycle, which is actually the process of  intelligent decision-making. When considering the CR networks, the cognitive devices are expected to have the perception capability, which enables the cognitive users to learn from the radio environment. Spectrum sensing actually provides cognitive users the capability to learn the spectrum holes for secondary access. Based on the wireless parameters learnt by the cognitive devices, intelligent decision will make the users adaptive to the radio environment. For example, the intelligent decision could  maximize the utility of spectrum resource and/or extend the lifetime of cognitive devices. After the cognitive devices are reconfigured based on the intelligent decision, the feedback, i.e., the influence of the decision, will be provided to the cognitive devices, which is also taken as the observations from the environment.

In fact, CR is preferable in a simple and predicable radio environment. With further evolution of cellular communications to the 5G and its beyond, future wireless networks become more complicated and unpredictable than ever before \cite{david20186g,patzold20195g,agiwal2016next}. As a result, CR faces the following challenges. First, with the exponential increase of communication devices (including both mobile devices and small base stations), the wireless networks in the 5G and beyond will be at large-scale with heterogeneous network typologies, which makes it costly for the CR devices to learn a complete and accurate radio environment information. Second, users may have various service demands (e.g., requests for text, audio, or video contents) with different technologies (e.g., 2G to 5G, and WiFi). In brief, the radio traffic models in the 5G and beyond are highly dynamic, which makes it difficult for CR devices to learn and predict. Third, with the visualization and cloudification of wireless networks in the 5G and beyond, multiple-dimensional resources (e.g., time, spectrum, spatial, computing, storage) at different layers (e.g., physical layer, link layer, network layer) need to be  coordinated and allocated. Typically, solving the multiple-dimensional resource allocation problem requires high  complexity to obtain the optimal or near-optimal solution.

Motivated by the outstanding performance,  \textit{artificial intelligence} (AI) technologies  have been applied in many areas recently, such as computer vision and wireless communications, and shown powerful learning capabilities in both simulation and in-field experiments~\cite{Jiang:2017:wm,Zhijin:WCM:2019}. The main difference between the intelligent decision-making processes with a cognitive agent and an AI agent is that, the AI agent is  more powerful in terms of learning capability:
\begin{itemize}
    \item First, the AI agent has a better generalization functionality than the cognitive agent. It can learn a robust pattern of the environment and make a proper action decision even though it has incomplete and inaccurate information of the environment in a large-scale and heterogeneous network.
    \item Second, the AI agent has a better predictable functionality than the cognitive agent. Therefore, it can track the variation pattern of the radio environment and infer a proper action decision in a highly dynamic wireless network.
    \item Third, the AI agent has a better reasoning functionality compared to the cognitive agent. As a result, it can avoid complicated mathematical formulations and therefore can learn the impact of an action on the environment quickly. Moreover, the AI agent can make the optimal or near-optimal action decision in a prompt manner.
\end{itemize}

There have been some excellent  survey and tutorial articles on CR in the past 20 years. \cite{Mitola:1999}  first introduces the concept of CR and discusses its relationship with \textit{software defined radio} (SDR), which provides important insights for the implementation of CR and SDR. This can be regarded as the beginning of  the prosperous period of CR. Lately, for the first time, \emph{Haykin et al.} \cite{Haykin:JSAC:2005} have discussed the basic CR functionalities from communications, signal processing, and networking point of view. Moreover, they have introduced the methods for radio scene analysis, channel state and interference-temperature estimation, and power control in CR. Additionally, \emph{Zhao et al.}~\cite{Zhao:SPM:2007} have unified the terminology of CR and \textit{dynamic spectrum access} (DSA), and provided an overview on challenges and recent developments in both technological and regulatory aspects of DSA. The xG network architecture, including spectrum management, spectrum mobility, and spectrum sharing, has been explained in~\cite{Akyildiz:2006:NGS:1162469.1162470}. \emph{Goldsmith et al.}~\cite{Goldsmith:Proc:2009} have surveyed  CR networks in terms of information-theoretic capacity results, related bounds, and the degrees of freedom for different design paradigms, such as  underlay, overlay, and interweave paradigms. As one of the core enablers of CR, spectrum sensing has been reviewed extensively afterwards~\cite{Ma:Proc:2009,Yucek:CST:2009,Zeng2010,Liang:TVT:2011,Axell:May:2012}. Later on, \cite{Lu2012} has provided a summary for the first ten years' achievements on spectrum sensing and sharing in CR networks.

By introducing machine learning  to CR or  wireless communication systems, intelligent communications will be with might doubled~\cite{ZQ:IR:2019}.  As we will see in this article, machine learning can significantly improve the performance of physical layer processing and MAC layer in communications. More importantly, intelligent communications can deal with some complicated tasks that traditional communications are unable to. The Internet of things (IoT), vehicular communications (V2X), and UAV based communications are three important application scenarios of future wireless networks. Only with machine learning, the complicated issues, such as resource allocation and routing in IoT and V2X, trajectory optimization in unmanned aerial vehicle (UAV) based communication networks, can be well addressed. Therefore, intelligent communications are the future trend for wireless networks to satisfy various demands of different applications. With the recent boom on AI and its applications in wireless communications, a comprehensive review on the evolution from cognition to intelligent communications is more than desired.

This article will provide an overview on the remarkable achievements in the area during the past 20 years. We aim to provide a big picture of the development of wireless communications from cognition to AI. The rest of this article is organized as follows. Section~\ref{Sensing} provides an overview on the  machines' perception ability with particular focus on typical and AI-enabled spectrum sensing in intelligent communications. Section~\ref{Sharing} reviews the machines' action of the perception-action cycle, which refers to the interaction between cognitive devices with the wireless environment. Particularly, typical and AI-enabled spectrum sharing in intelligent communications will be demonstrated respectively. Section~\ref{IC} identifies the research challenges that should be addressed in the future before the realization of AI-enabled intelligent communications. Section~\ref{Conclusions} concludes this article.

\section{Perception}\label{Sensing}
As aforementioned, the perception capability from radio environment is one of the key components in intelligent communications. In CR networks, the perception process mainly focuses on identifying vacant channels for secondary users to access. Moreover, parameters, such as \textit{channel state information} (CSI), interference, and modulation type,  could also be learned to facilitate  intelligent decision making. In the following of this section, we start from the traditional spectrum sensing techniques with highlighting the remarkable work during the past 20 years. Then we will provide an overview on the recent achievements in intelligent communications as inspired by the boom of AI.

\subsection{Traditional Spectrum Sensing}
\subsubsection{Narrowband Spectrum Sensing}
In CR networks, spectrum sensing is regarded as one of the most challenging tasks. By performing spectrum sensing, secondary users will have the knowledge of spectrum occupancy. Once a spectrum hole is identified, secondary users can use it for data transmission. There has been extensive research work on spectrum sensing, which mainly includes matched filter detection, cyclostationary feature detection, and energy detection. The matched filter detection is an optimal detection method that requires the prior information of primary users. However, it requires secondary users to equip with a dedicated sensing receiver for each type of primary signals. The benefit of adopting cyclostationary feature detection is that it is able to distinguish the primary users and noise by utilizing the periodicity in the received primary signal. However, high computational complexity and prior information of the primary signals are normally required. Energy detection is a non-coherent detection method, which can avoid the requirement for prior knowledge of primary users. Moreover, energy detection alleviates the requirement for complicated receivers while the other two approaches normally need complex receiver design. Therefore, energy detector is easier to be implemented and the complexity is usually lower, but its detection performance is poor under low \textit{signal-to-noise ratio} (SNR) scenarios.

In the past 20 years, a large group of engineers and researchers have made great efforts to spectrum sensing, in which the probability of detection and the probability of false alarm are normally taken as two  performance metrics. With higher  detection probability, the primary users can be protected better. However, from the secondary users' perspective, with  lower  false alarm probability, spectrum resource can be reused at a higher probability when it is available. Therefore, higher throughput can be achieved by the secondary networks. Take the simplest energy detector as an example, the threshold for spectrum occupancy determination is dependent on the size of sampling samples and the SNR. \cite{Tandra:JSAC:2008}  models the effects of noise and channel fading uncertainty, which could be quantified by the term ``SNR wall''. Particularly, the tradeoff between the capacity of primary users and the sensing robustness of secondary users has been quantified for some simple detectors. It has been pointed out that below the SNR wall, a detector fails to be robust regardless of the sensing period.  Meanwhile, another pioneer work~\cite{Liang:TWC:2008} firstly attempted to  optimize the sensing duration  to maximize the achievable throughput for the secondary networks while providing sufficient protection to the  primary users. This inspiring work has started the rapid development of spectrum sensing in terms of system throughput optimization with various constraints.

For the aforementioned  spectrum sensing techniques, the sensing performance  is often affected by interference, noise, and fading of wireless channels. Inspired by the cooperative diversity~\cite{Laneman:2004}, cooperative spectrum sensing~\cite{Haykin:JSAC:2005,Ghasemi:DYSPN:2005,Ganesan:2005,Ganesan:2007:twc,Ganesan:2007} has been proposed to exploit observations or data from multiple CR users to improve sensing performance.  Various cooperative sensing techniques have been developed afterwards, which can be categorized into two types,  the centralized ones and the decentralized ones.

In the centralized cooperative sensing, multiple CR users send observed or processed data on the sensed spectrum to the fusion center, which is normally powerful for data processing. The fusion center then combines all observations from different CR users and makes a decision on the spectrum occupancy. Since sending observations to the fusion center costs spectrum and power resources, the format and amount of observations to be sent to the fusion center are dependent on the available resources and the specific data combining and detection method. Particularly, multitaper spectral estimation combined with singular value decomposition has been introduced for cooperative spectrum sensing, which requires soft-data or soft-decision from the cooperated CR users~\cite{Haykin:JSAC:2005}. To save the spectrum resource allocated for report channels and reduce the computational complexity at the fusion center, only the hard-decisions are required from the cooperated CR users~\cite{Ma:TWC:2008}. Correspondingly, the sensing performance is degraded. In general, it is a trade-off between the complexity and sensing performance.

In the decentralized cooperative sensing, a CR user can get data from the other cooperative CR users, which are usually nearby, through relays or \textit{device-to-device} (D2D) communications. The CR user may act as the fusion center and use the decision methods similar to those adopted in the centralized approaches. Some methods tailored for decentralized cooperative sensing have been also developed in~\cite{Ghasemi:DYSPN:2005,Ganesan:2005,Ganesan:2007:twc,Ganesan:2007}. More information on cooperative spectrum sensing can be also found in~\cite{Ma:Proc:2009,Letaief:PROC:2008,Liang:TVT:2011,Zhou:2018} and the references therein.

\subsubsection{Wideband Spectrum Sensing}
The aforementioned work mainly focuses on the narrowband sensing, which normally implies that the frequency range is sufficiently narrow such that the channel frequency response can be considered as flat. In other words, the bandwidth of interest is less than the coherence bandwidth of the channel. To find spectrum holes for opportunistic access, the secondary user scans the channels of interest one by one until the vacant one is identified. It is more efficient for spectrum discovery if the secondary user could sense multiple channels simultaneously. However, the narrowband sensing techniques cannot be applied to wideband spectrum sensing straightforwardly, as the narrowband techniques normally make a  binary decision for the whole spectrum, which cannot identify the individual spectral opportunities that lie within the wideband spectrum~\cite{sun_wideband:2013}. Therefore, secondary users are expected to exploit spectral opportunities over a wide frequency range and to identify multiple spectrum holes within one sensing period.

A straightforward approach of performing wideband spectrum sensing is to acquire the wideband signal by a high-speed \textit{analog-to-digital converter} (ADC), and then digital signal processing techniques are utilized to detect spectral holes. Some research on wideband spectrum sensing have been carried out  with the implementation of a high-speed ADC. A typical approach is to perform sampling over the wideband signal by a high-speed ADC. Subsequently, the received signal is processed by a serial-to-parallel conversion circuit to divide sampled data into parallel data streams. Additionally, \textit{fast Fourier transform} (FFT) is implemented to convert the wideband signals from the time domain to the frequency domain. As a result, the wideband spectrum signal is divided into series of narrowband ones. The energy of each one is then calculated by adopting energy detector. Finally, the spectrum occupancy of each narrowband channel is determined by using an optimized threshold to achieve better detection performance than narrowband spectrum sensing approaches~\cite{Quan:JSTSP:2008}.

According to the Nyquist-Shannon sampling theory, the sampling rate must be no less than twice of the maximum frequency presented in the signal (known as Nyquist rate), in order to avoid spectral aliasing. However, such high-speed ADCs are unaffordable for most of the CR devices. Therefore, wideband spectrum sensing presents significant challenges on hardware to operate at very high sampling rates. With current hardware technologies, high-rate ADCs with high resolution and reasonable power consumption (e.g., 20 GHz sampling rate with 16 bits resolution) are difficult to design. Even if it were possible, the real-time digital signal processing of high-rate sampling could be very expensive.

\begin{figure}[!t]
           \centering
            \includegraphics[scale=0.62]{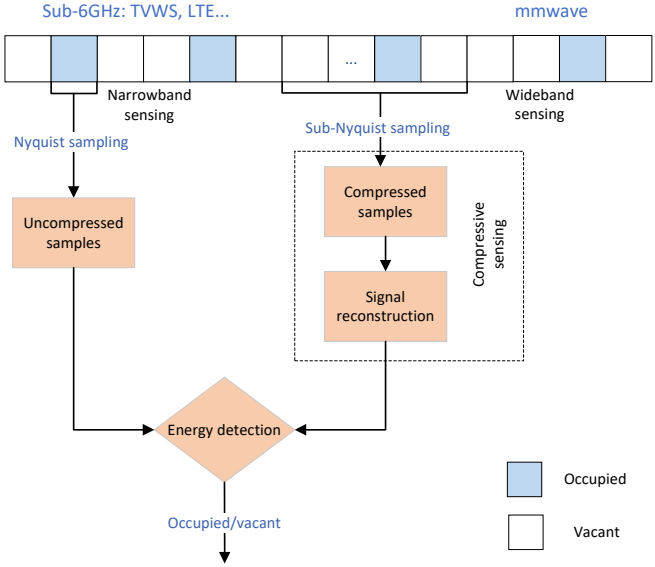}
            \caption{Comparison of narrowband and wideband spectrum sensing based on compressive sensing.}
           \label{CSS}
\end{figure}

With the help of its most recent development on compressive sensing techniques, the bottleneck of wideband spectrum sensing can be broken by utilizing the sparsity property of spectrum~\cite{QIN:SPM:2018}. Compressive sensing has been firstly introduced in~\cite{zhitian:2007}, which enables sub-Nyquist sampling over a wide frequency band  without loss of any information. Fig.~\ref{CSS} shows the basic principle of compressive  spectrum sensing in comparison with the typical narrowband spectrum sensing. The core idea of compressive sensing based wideband spectrum sensing, namely compressive spectrum sensing, is to shift the burden on high-speed ADCs to the digital signal processing after sampling.  In order to realize sub-Nyquist sampling over a wideband channel, extensive research work~\cite{Meng:JSAC:2011,Zhijin:TSP:2015,QIN:TCOM:2017} has been carried for both the single CR device case and the cooperative networks with multiple CR devices by following the research route in the narrowband spectrum sensing.

Compressive spectrum sensing can be categorized into two types. The first type exploits the sparsity of spectral signals in the frequency domain caused by the low spectrum utilization. The sparsity level needs to be estimated first to determine the minimal sampling rate at secondary users~\cite{yue_sparsity:2012,QIN:TWC:2016,QIN:TSP:2018}. A two-step CS scheme has been proposed in~\cite{yue_sparsity:2012}  to minimize the sampling rates when the sparsity level is changing. However, introducing an extra step on the sparsity estimation could be expensive or even unaffordable for cognitive devices. The prior information from geo-location database has been  utilized to provide a rough estimation of the sparsity~\cite{QIN:TWC:2016}, and to recover  signals  with reduced complexity and improved accuracy. Moreover, this type of approaches may fail if the spectrum utilization is quite high or the noise level is very high. Another type of compressive spectrum sensing utilizes the cyclic feature~\cite{Tian:JSTSP:2012,Cohen:2017}, which inherits the robustness to noise of cyclic spectrum sensing techniques. Tian \emph{et al.}~\cite{Tian:JSTSP:2012} have used time-varying cross-correlation functions of compressed samples to get the cyclic spectrum. It is also able to  recover the power spectra of stationary signals, which makes the approach applicable even for non-sparse signals. It has been further proved that cyclic spectrum can be reconstructed from sub-Nyquist samples without sparsity constraint on the signals~\cite{Cohen:2017}.

\subsubsection{Other Perception Applications}
Apart from the aforementioned work on spectrum sensing, CR devices' perception capability also involves the modulation and waveform design, as well as the propagation modeling. The whole idea behind CR is that enough protection should be guaranteed for primary users. Therefore, the mutual interference between secondary users and primary users should be controlled to enable their coexistence. The modulation strategies have been reviewed in~\cite{Budiarjo:SPM:2008}, which provides  an excellent exposition to the \textit{orthogonal frequency-division multiplexing} (OFDM) and \textit{transform domain communications system} (TDCS) modulation techniques for spectrum overlay-based CR systems. Moreover, the propagation channel model should be considered carefully to enable the implementation of CR, which could be deployed over a wide range of the spectrum, including the \emph{ultra high frequency} (UHF) bands, cellular bands, and fixed wireless access bands. The \textit{millimeter wave} (mm-wave) provides more spectrum resources for opportunistic access. Over these bands, wireless signals are normally affected by the propagation in different ways. \emph{Molisch et al.}~\cite{Molisch:2009} have provided a comprehensive overview on the propagation channel characteristics and models, which are essential for the design of spectrum sensing methods and transmission strategies for CR systems.

\subsection{Learning from Radio Environment}

In recent years, machine learning techniques are widely applied to learn the radio environment. The  motivation of applying machine learning algorithms in wireless communications is that, historical wireless data contains the features and variation patterns of the radio environment, which can be used for parameter configuration and performance enhancement. The promising machine learning algorithms can be divided into three categories: supervised learning, unsupervised learning, and reinforcement learning (RL). In this part, we will first provide brief introductions of typical machine learning algorithms followed by their applications in the radio environment learning.

\subsubsection{Supervised learning}
The agent in supervised learning learns features and patterns hidden in the labeled data. If we denote $x_n$ as the $n$-th input and  $y_n$ as the corresponding output, we can define $(x_n, y_n)$ as the $n$-th label. The mapping between the input $x_n$ and the output $y_n$, $f(\cdot): x_n \ \to \ y_n$,  can be interpreted as the impact of the radio environment (e.g., channel, interference, and noise) on the input. The goal of the agent is to learn $f(\cdot)$ from labels and  infer the output for any input in the future based on the learnt $f(\cdot)$. 

In general, applying supervised learning in the perception of radio environment consists of two steps. First, historical wireless data can be separated into the radio environment data set (i.e., inputs of the supervised learning) and the action data set (i.e., outputs of the supervised learning). In particular,  the radio environment data set is divided into different groups, each of which is labelled with a unique action. By doing so, historical wireless data can be used as the labelled data for supervised learning. Second, by using the labelled historical wireless data and adopting proper models to learn the mapping between the radio environment data and the action, the agent is able to learn the interaction relationship between the radio environment data and the action data. Then, the agent can make a proper action decision after the new environment data arrives to the learned mapping. Typical models that can be used to learn the mapping between the radio environment data and the action data include \textit{$k$ nearest neighbours} (KNN), \textit{support vector machine} (SVM), and \textit{artificial neural networks} (ANN).

KNN is one of the simplest model in supervised learning, in which the data points with similar profiles are generally in close proximity according to a certain distance metric, regardless of the distribution of the data points. In the KNN enabled perception of wireless radio environment, historical radio environment data can be firstly categorized into different groups, each of which is labelled with an unique action. By clustering new radio environment data into a proper group, the relationship between the new radio environment and the expected action is considered to be similar to that between the radio environment data in the group and the labelled action. Then, the labelled action can be directly adopted for the new radio environment. In fact, the radio environment is a broad concept in wireless communications. Depending on different kinds of purposes, only the related radio environment data should be used, which is usually selected through a trial and error manner. For example, the number of users, CSI, and interference level are related to the beam selection scheme~\cite{KNN_1}. The power strength of the received primary signal samples can indicate spectrum occupancy and thus can be used as the core radio environment data for the spectrum detection in CR networks \cite{KNN_2}. Besides, the signal strengths received at a specific receiver from unknown transmitters distributed over distinct locations are quite different, which enables them to be used as key radio environment data for localization \cite{KNN_3}.

The KNN model is applicable when the radio environment data is linearly separable. If the radio environment data is not linearly separable in its original space, SVM is a good alternative, which adopts kernel functions to map the data from its original space to a higher-dimension space, such that these data become linearly separable in the new space. In \cite{SVM_1}, the spectrum hole data including frequency feature, power feature, and time feature have been used to identify different \textit{medium access control} (MAC) protocols used by received signals. As these data are linearly inseparable in its original space, the SVM model has been adopted to cluster the spectrum hole data into different groups, each of which corresponds to a unique MAC protocol. Similar ideas have been adopted in \cite{SVM_2,SVM_3,SVM_4} for modulation classification and spectrum detection.

As aforementioned, the  radio environment data used in both KNN model and SVM model is manually designed through a trial and error manner. With the expansion of wireless systems, the best action may be a complicated composition function of various types of radio environment data, which makes the system being overwhelmed by  massive raw  data of radio environment. Therefore, it is challenging for the agent to identify the related radio environment data among the massive raw  data, which is helpful to make the best action decision. Moreover, the relationship between the used radio environment data and the best action for the specific purpose may be complicated and the performance of the KNN model and the SVM model are unsatisfactory. To deal with the issues, convolutional neural network (CNN) is usually adopted due to its powerful representation learning capability \cite{Zhijin:WCM:2019}. In particular, a CNN model is generally composed of a convolution part and a classification part. The convolution part  automatically extracts the main features of raw  radio environment  data and the classification part approximates the complicated functions (i.e., mapping relationship) between the extracted features and the best actions. \cite{lee2019deep} has adopted the CNN model for cooperative spectrum detection. By using the raw   primary signal strengths received at multiple detectors as the inputs, the CNN model can learn a better mapping relationship between the raw data  and the  detection results as well as achieve a better spectrum detection performance compared with the SVM model. Other applications of deep learning for the perception include joint channel estimation and signal detection~\cite{Ye:2018}, link adaption \cite{elwekeil2018deep},  waveform recognition \cite{zhang2017convolutional}, and radio localization \cite{bregar2018improving}.

\begin{figure}[!t]
    \centering
    \includegraphics[scale=0.5]{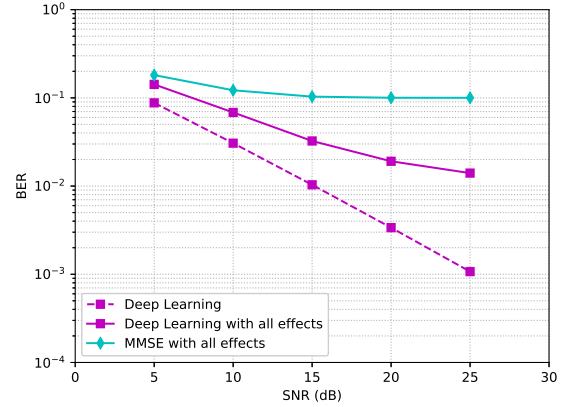}
    \caption{Performance comparison of channel estimation and signal detection based on deep learning and MMSE methods.}
    \label{DL_CSD}
\end{figure}

Fig. \ref{DL_CSD}, from~\cite{Ye:2018},  shows the  performance when the deep learning technique is used for the joint channel estimation and signal detection~\cite{Ye:2018}. In the simulated scenario, the cyclic prefix is omitted and the clipping noise is considered. It can be observed that DNN significantly outperforms the minimum mean square error (MMSE) based approach in terms of bit-error rate.

\subsubsection{Unsupervised learning}
The agent in unsupervised learning learns the main features and patterns hidden in the unlabeled data. Typical applications of unsupervised learning include clustering and data compression.

Unsupervised clustering algorithms can be roughly classified into model-based clustering and model-free clustering. In particular, model-based clustering usually assumes that the data in each group follows a certain distribution, while there is no assumption on the data distribution  in the model-free clustering. One typical application of model-based clustering for radio environment perception is symbol detection \cite{zhang2018label,jilkov2010design,yang2015unknown}. For example, in \cite{zhang2018label}, by using Gaussian distribution to model the received symbols in each group that corresponds to a unique transmitted symbol, the received symbols in multiple groups follow the \textit{Gaussian mixture model} (GMM) and the parameters, e.g., mean and variance, of the Gaussian distribution can reflect the impacts of the radio environment on the transmitted symbols. In other words, by adopting GMM-based clustering, the agent can estimate the parameters of the Gaussian distribution and learn the mapping relationship from the transmitted symbols and the received symbols. Then, each received symbol can be properly clustered and detected. One typical application of model-free clustering for radio environment perception is communication node (BS or user) clustering. For example, in a large-scale network with multi-dimensional radio resource, it is challenging to  manage the radio resource in a centralized manner due to high signalling costs. Alternatively, nearby communication nodes can be clustered into the same group, in which the radio resource can be coordinated among the communication nodes \cite{samarakoon2016dynamic,zhao2015using,echoukairi2017novel,zhou2015distributed}. The intuition behind is two-fold. On one hand, it is easy to exchange signalling among nearby communication nodes. On the other hand, nearby communication nodes have the highest impacts on the radio resource management for each other.

\textit{Principal component analysis} (PCA) is a typical data compression algorithm \cite{jolliffe2011principal}, which is usually used to extract expected signals from noisy signals or circumvent the multi-path impacts in wireless communications. In \cite{nasser2016spectrum}, the authors have proposed a PCA-based spectrum detection framework for CR networks. In particular, if the signal samples only contain white noise, the covariance matrix of the signal samples is diagonal. If the signal samples contain both primary signal and white noise, the covariance matrix of the signal samples can be represented as the summation of the diagonal matrix and a low-rank matrix since the covariance matrix of primary signal samples is usually low-rank. By subtracting the covariance matrix of the white noise samples from the covariance matrix of the signal samples and applying PCA into the remaining covariance matrix, the cognitive user can obtain the largest principal component of the remaining covariance matrix, which can be used as a good test statistic for spectrum sensing. In \cite{yoo2017indoor}, the authors have developed a PCA-based radio localization scheme. In particular, the \textit{received signal strength} (RSS) contains the information of user's location, which is a random variable due to  multi-path impacts. Then, PCA is adopted to analyze RSS samples and extract the location information.

\subsubsection{Reinforcement learning}
RL mimics the learning process in the brain via trial and error with purpose of finding the optimal action policy, which maximizes a long-term reward by continuously interacting with the environment \cite{sutton1998introduction}. Two representative RL algorithms are Q-learning and \textit{deep reinforcement learning} (DRL).

\begin{figure}[!tp]
          \centering
           \includegraphics[scale=0.65]{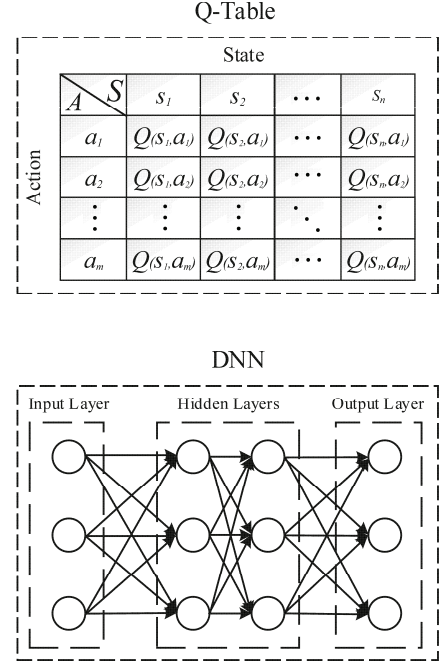}
           \caption{Illustrations of Q-table and DNN structure.}
          \label{RL}
\end{figure}

In the Q-learning, a Q-table is established with Q-values as elements as shown in the top of Fig. \ref{RL}. Here, each Q-value is defined as the discounted accumulative reward (long-term reward) for an arbitrary state-action pair, which indicates the impacts of the action on the state. By iteratively updating the Q-value of each state-action pair until convergence, the optimal action has the maximum Q-value and can be selected for executing. In fact, the Q-values update procedure is the perception that learns the mapping relationship from each state to its best action. In recent years, we have witnessed the wide applications of Q-learning for perception in wireless communications. For example,  Q-learning has been adopted to  find spectrum holes over wideband spectrum by considering the realistic hardware reconfiguration  and delays~\cite{li2014learning}. By doing so, the requirement on the complete knowledge of radio environment could be lowered. \textit{Biggelaar et al.}~\cite{van2012sensing} have proposed a distributed Q-learning algorithm to share the sensing time among cooperative users to maximize the throughput of the CR networks. Moreover, a distributed Q-learning algorithm has been designed to optimize the transmit power of CR users with the purpose of maximizing the \emph{signal to interference plus noise ratio} (SINR) at the secondary receivers while meeting the primary protection constraint.  It is worth noting that when the state-action space is relatively large, the performance of the Q-learning enabled perception drops since many state-action pairs may not be explored by the agent. Moreover, when the state-action space becomes infinite, the Q-learning algorithm is no longer applicable since it is impractical to establish an infinite Q-table.

To overcome the drawbacks of Q-learning algorithms, DRL adopts a deep NN (DNN), as shown in the bottom of Fig.~\ref{RL}, to approximate the Q-values. Particularly, when the state-action space is relatively large and the agent fails to explore all the state-action pairs, the DNN can still take proper actions for the states that have not been explored by the agent due to the good generalization property of the DNN. Moreover, the DRL agent only stores weights of the DNN instead of an infinite number of Q-values. Accordingly, the  weights update procedure  in the DNN is the perception procedure to learn the mapping  from each state to the corresponding best action. For instance, an intelligent modulation and coding selection~\cite{zhanglin2019deep}  has been developed  for the primary transmission where a DRL agent is implemented at the primary transmitter to  learn the interference pattern from secondary transmitters.  Moreover, \textit{He et al.} have used DNN to learn the impact of user scheduling on the sum-rate in a wireless caching network~\cite{he2017deep}. It is noted that the perception and action is usually coupled, especially when we adopt reinforcement learning techniques. More application of DRL for the perception and action in wireless communications will be detailed in Section III.

able~\ref{Comparison_table} provides a brief summary of different ML algorithms and their applications in  intelligent communication systems reviewed in this article.

\begin{table*}[!t]
\caption{Comparison of different machine learning algorithms and their applications in intelligent communications.}
\label{Comparison_table}
\begin{tabular}{|l|l|l|l|l|}
\hline
ML category                                                                       & Algorithms & Scenarios                                                                                                                               & Feature                                                                                                                       & Cons                                                                                                        \\ \hline
\multirow{3}{*}{\begin{tabular}[c]{@{}l@{}}Supervised\\ learning\end{tabular}}    & KNN        & \begin{tabular}[c]{@{}l@{}}Spectrum detection\\ Localization\end{tabular}                                                               & \begin{tabular}[c]{@{}l@{}}Applicable for linearly \\ separable data\\ One to one mapping\end{tabular}                         & \multirow{2}{*}{Require data labeling}                                                                      \\ \cline{2-4}
                                                                                  & SVM        & \begin{tabular}[c]{@{}l@{}}Spectrum detection\\ Modulation classification\end{tabular}                                                  & \begin{tabular}[c]{@{}l@{}}Applicable for linearly \\ non-separable data\\  One to one mapping\end{tabular}                     &                                                                                                             \\ \cline{2-5}
                                                                                  & ANN        & \begin{tabular}[c]{@{}l@{}}Spectrum detection\\ Channel estimation \& signal detection\\ Waveform recognition\\ Localization\end{tabular} & \begin{tabular}[c]{@{}l@{}}Provide better mapping\\ between raw data and\\ action compared to \\ KNN and SVM\end{tabular}     & \begin{tabular}[c]{@{}l@{}}Require data labeling\\ Overfitting problem\end{tabular}                         \\ \hline
\multirow{2}{*}{\begin{tabular}[c]{@{}l@{}}Unsupervised\\ learning\end{tabular}}  & GMM        & \begin{tabular}[c]{@{}l@{}}Symbol detection\\ Communication node clustering\end{tabular}                                               & \begin{tabular}[c]{@{}l@{}}Mainly for clustering \\ problem\end{tabular}                                                      & \begin{tabular}[c]{@{}l@{}}Require prior knowledge\\ of distribution of \\ received symbols\end{tabular}    \\ \cline{2-5}
                                                                                  & PCA        & \begin{tabular}[c]{@{}l@{}}Spectrum detection\\ Localization\end{tabular}                                                               & \begin{tabular}[c]{@{}l@{}}Mainly for data \\ compression\end{tabular}                                                        & \begin{tabular}[c]{@{}l@{}}Information loss in\\ original signal\end{tabular}                               \\ \hline
\multirow{2}{*}{\begin{tabular}[c]{@{}l@{}}Reinforcement\\ learning\end{tabular}} & Q-learning & \begin{tabular}[c]{@{}l@{}}Wide spectrum sensing\\ Spectrum sharing\\ System parameter reconfiguration\end{tabular}                    & \begin{tabular}[c]{@{}l@{}}Q-table to store the \\ relationship between \\ state and action\\ One to one mapping\end{tabular} & \begin{tabular}[c]{@{}l@{}}Infinite state-action\\ space leads to \\ unaffordable\\ complexity\end{tabular} \\ \cline{2-5}
                                                                                  & DRL        & \begin{tabular}[c]{@{}l@{}}Spectrum sharing\\ User scheduling\\ System parameter reconfiguration\end{tabular}                          & \begin{tabular}[c]{@{}l@{}}Applicable to problems with \\ infinite Q-table\\ Only store weights\end{tabular}                    & \begin{tabular}[c]{@{}l@{}}Complex training for \\ multi-agent case\end{tabular}                            \\ \hline
\end{tabular}
\end{table*}

\section{Action}\label{Sharing}
Based on the wireless environment features learnt by the cognitive users, intelligent decisions can be made so that the devices can be reconfigured to adapt to the radio environment and maximize the utility of the radio spectrum resource.

\subsection{Traditional Spectrum Resource Allocation and Sharing}
The traditional resource allocation and sharing schemes in CR can be categorized based on four different access paradigms, namely interweave, underlay, overlay, and hybrid communications, which will be detailed in the following.

\subsubsection[Interweave]{Interweave\footnote{Interweave is referred to as overlay in some literature.}}

Secondary users can exploit spectrum holes, i.e., gaps in time, frequency, and space that are not occupied by primary users in the interleave paradigm. Obviously, the capability of perception from radio environment is very important to identify the spectrum holes for the secondary users to communicate in an opportunistic manner. The aforementioned perception techniques, such as spectrum sensing, are therefore essential to interweave communications. The more reliable the perception is, the less interference between the primary and secondary users will be generated.

OFDM and \emph{orthogonal frequency-division multiple access} (OFDMA) are attractive transmission and multiple access techniques \cite{Weiss04} for interweave communications, given their flexibility in turning on or off tones and utilizing non-adjacent sub-bands to adapt to spectrum holes in the radio environment. However, even with perfect spectrum sensing, \emph{out-of-band} (OOB) leakage of the OFDM signal would still bring interference to the primary and secondary users.
In \cite{4657096}, the joint subchannel, rate, and power allocation for secondary users sharing frequency bands with primary users using OFDM has been considered and an optimization problem to achieve max-min rate sharing among the users has been formulated. Both optimal and suboptimal approaches are proposed and compared.
The problem of subcarrier and power allocation in multicast wireless systems using OFDMA has been studied in \cite{5164968}. To maximize the system throughput and ensure minimum numbers of subcarriers for individual multicast groups, low-complexity schemes have been proposed by separating subcarrier and power allocations and with a modified genetic algorithm, respectively.
With the consideration of the interference constraint to primary users and the upper and lower bounds on the bandwidth for individual secondary users, the joint subcarrier and power allocation in OFDMA-based ad hoc CR networks has been addressed in \cite{5776714}. Distributed protocols with the use of a common reserved channel have been proposed to reduce the computational complexity while attaining the optimality of the solution.
Moreover, resource allocation for wireless virtualization to assign the physical spectrum resources to several virtual networks has been considered in \cite{6603649} and the problem of resource allocation with carrier aggregation for spectrum sharing between a \emph{multiple-input multiple-output} (MIMO) radar and a \emph{Long Term Evolution  Advanced} (LTE-A) cellular system has been studied in \cite{6817777}.

\subsubsection{Underlay} Secondary users may transmit over the same frequency band and at the same time as primary users in the underlay paradigm. However, the interference from the secondary transmitters to the primary receivers must be   controlled carefully. In underlay communications, the tolerable interference level at a primary receiver can be modeled by the interference temperature concept defined by the Federal Communications Commission (FCC) \cite{federal2003establishment}. To ensure the reliable operation of the primary users, the interference constraint can be very restrictive. As a result, the secondary transmitters are typically very conservative in their transmit powers.

The problem of resource allocation for underlay communications has been discussed in \cite{4723340}, where both the interference tolerance for primary users and \emph{quality-of-service} (QoS) requirement that translates to SINR for secondary users are taken into account. Admission control algorithms have been proposed together with power control to satisfy the constraints for both primary and secondary users. Meanwhile, optimization problems for rate and power allocation under proportional and max-min fairness criteria have been formulated and solved. In \cite{5581200}, the rate and power adaptation in spectrum sharing to maximize the achievable capacity of the secondary user with interference power constraints and bit-error-rate requirements have been considered. The benefits of soft-sensing information on primary user activity are shown in different operating scenarios. Based on the available channel state information and the constraint for spectrum sharing, two lower bounds of the mean rate for the primary user have been derived in \cite{5699937}. With only the secondary-to-secondary and secondary-to-primary link gains, a power control policy is then proposed to guarantee minimum instantaneous rates for both the primary and secondary users. Similarly, the problem of admission and power control has been studied in \cite{5751194}, with a strict interference power limit and a minimum QoS requirement. It is shown that the problem to maximize the number of admitted secondary links or the sum rate of the admitted secondary links is either NP-hard or non-convex; therefore, suboptimal algorithms have been proposed. With only partial CSI, the resource allocation problem in OFDMA-based spectrum sharing systems has been studied in \cite{5967979} to maximize the weighted sum rate of secondary links given the service collision probability constraint for primary links. As the original optimization problem is non-convex, dual optimization method has been used to obtain suboptimal solutions with reduced complexity. Moreover, a distributed pricing scheme has been proposed in \cite{1542621}, where users exchange ``price" signals to indicate the negative effect of interference at the receivers.
As a result, each transmitter can choose a channel and power level to maximize its net benefit, i.e., utility minus cost. The proposed pricing algorithm outperforms the heuristic algorithm and may outperform the iterative water-filling algorithm in a dense network.

\subsubsection{Overlay} In the overlay paradigm, secondary users may also transmit over the same frequency band and at the same time as primary users. Different from the underlay communications, the restrictive transmit power limit is lifted in overlay communications. To offset the interference generated by a secondary transmitter at a primary receiver, a portion of the transmit power of the secondary user is used to assist the transmission of the primary user. Therefore, the overlay paradigm requires cooperation between the primary and secondary users so that the secondary system has certain knowledge about the primary system and uses it to design advanced coding and transmission schemes.

For example, when the primary system is unable to achieve the target transmission rate, the secondary system acts as a relay and helps the primary system to forward the primary signal with a fraction of the subcarriers. Meanwhile, the secondary system uses the rest of the subcarriers to transmit its own signal as in opportunistic spectrum access. In \cite{6177985}, the optimization of the set of subcarriers allocated for cooperation and power allocation are considered to maximize the transmission rate of the secondary system while allowing the primary system to achieve the target rate.

\subsubsection{Hybrid} To overcome the drawbacks of the above paradigms, the hybrid paradigm \cite{sharma2014hybrid,jiang2013hybrid} mixes some of the above paradigms. For example, the interweave paradigm does not consider the tolerable interference level at a primary receiver while the underlay paradigm does not allow secondary transmission at a full power level. In contrast, a hybrid scheme may enable a secondary user to access an occupied frequency band with a controlled power and an idle frequency band with a full power. This paradigm has received great attention in the recent literature even though the term ``hybrid" is not always explicitly used.

In \cite{6392312}, a hybrid overlay/underlay spectrum sharing scheme has been employed, where the secondary users adapt its access to the licensed spectrum based on the status of the primary user. If the licensed spectrum is detected to be idle, the secondary user operates in the overlay mode. Otherwise, it selects the underlay mode. When there are multiple secondary users, an auction-based power allocation scheme is proposed so that the power can be allocated based on the payment of the secondary user and QoS of the primary user.

\subsubsection*{Resource allocation and sharing in heterogeneous networks and D2D communications}

The deployment of femtocells is considered as a promising solution to enhance the indoor coverage and the network capacity. Conventionally, the spectrum allocated to femtocells is from the same licensed bands of macrocells. Given the limited number of licensed spectrum bands, spectrum sharing between the macrocells and femtocells becomes necessary and the interference between macrocells and femtocells must be carefully managed. In \cite{5620930}, CR is incorporated into femtocell networks so that the femtocells can access spectrum bands not only from macrocells but also from other licensed systems. Different from traditional spectrum sharing schemes, such as coloring, decomposition theories are used and shown to achieve extra capacity.
To mitigate the cross-tier interference that limits the system performance, resource allocation for co-channel femtocells has been considered in \cite{6825834}. The subchannel and power allocation problem has been modeled as a mixed-integer programming problem to maximize the capacity with QoS and interference constraints, which can be transformed into a convex optimization problem and solved via the dual decomposition method. Moreover, a low-complexity distributed algorithm has been developed.

D2D communications has been proposed to underlay cellular networks and allow direct transmissions between local devices, which is promising to enhance the spectrum utilization in LTE-A networks \cite{6560489,6807946}. Similar to the femtocells, D2D communications may cause interference to the primary cellular communications when these two types share the spectrum bands. If the radio resource can be allocated intelligently, the interference can be mitigated. In \cite{5645039}, the problem of resource allocation in D2D communications has been formulated as a mixed integer nonlinear programming problem, where an alternative greedy heuristic algorithm has been proposed to reduce the interference. Based on a pricing scheme, interference coordination for D2D communications has been discussed in \cite{6949132}. The admission control and power allocation for D2D communications with QoS requirement for both D2D and cellular users has been studied in \cite{6362527}, where a set-based admission control algorithm and a distributed power optimization algorithm have been proposed. In \cite{6547816}, a new spectrum sharing protocol, enlightened by the overlay paradigm, has been proposed so that bi-directional communications of the D2D users is enabled, which can assist the two-way communications between the cellular base station and the cellular user. The achievable rate region is discussed and the optimization of power control and relay selection renders further performance improvement. In \cite{6590055}, joint resource block scheduling and power control has been further proposed for D2D communications in LTE-A networks. With the newly introduced D2D communication mode in addition to the conventional cellular mode, mode selection and switching, together with resource allocation, have been discussed in \cite{6924793} and \cite{7174559}.

\begin{figure}[!t]
           \centering
            \includegraphics[scale=0.45]{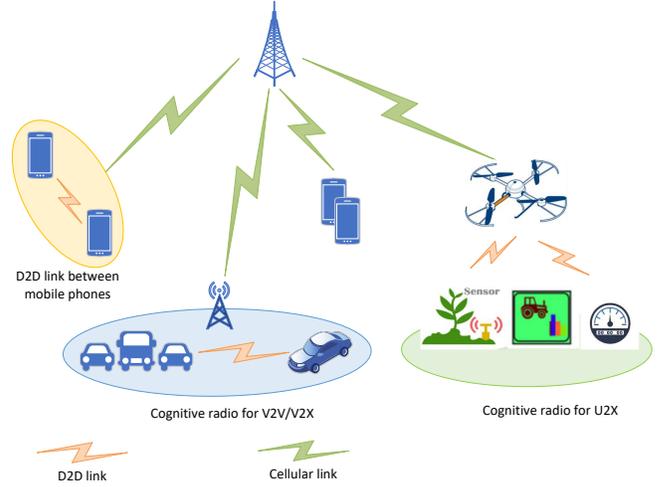}
            \caption{Applications of CR in various scenarios.}
           \label{CR_D2D}
\end{figure}

Fig.~\ref{CR_D2D} shows various application scenarios of CR, such as D2D, V2X/V2V, and UAVs to sensors (U2X). The resource allocation and sharing in D2D communications have been introduced above. Note that the resource allocation and sharing in V2X and drone communications  are more complicated. Therefore, the intelligent action should be taken, which will be discussed in next section.

\subsection{Intelligent Action}

\subsubsection{Reinforcement learning enabled intelligent action}

Machine learning, especially RL, is extremely useful for intelligent actions. When the wireless environment is learned by the decision maker, such as a cognitive user, it can adjust its operating parameters to adapt to the environment and maximize the system utility. However, the effect of operating parameters on the system utility is not always clear. Even with the perception capability, there always exists some uncertainty, where machine learning techniques can be applied to enhance the overall system utility. Among the categories of machine learning algorithms discussed above, RL would find plenty of opportunities in intelligent actions and utilizing the system resources.

When little knowledge is known about the effect of the operating parameters on the system utility, RL can use a stochastic finite state machine to model the wireless environment with inputs and outputs. The inputs can be the chosen operating parameters and the outputs can be the observations of the system utility for the cognitive user. To maximize the system utility, the environment is explored and then exploited.

\begin{figure}[!t]
           \centering
            \includegraphics[scale=0.55]{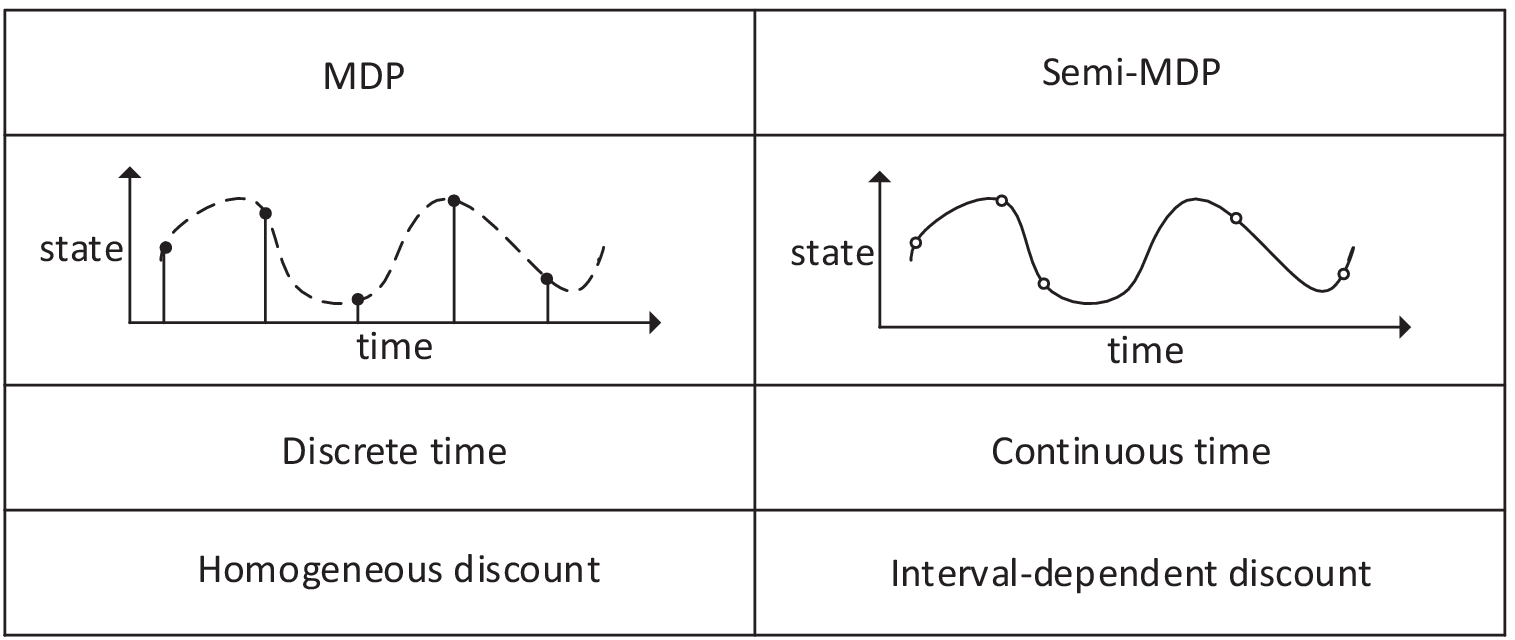}
            \caption{Comparison of MDP and semi-MDP.}
           \label{MDP}
\end{figure}

Note that an \emph{Markov decision process} (MDP) can be used to model decision-making under uncertainty, which is usually characterized by a tuple of four components ($S$, $A$, $T$, $R$), where $S$ is the state space, $A$ is the action space, $T(s, a, s^{'})$ is the probability of reaching state $s^{'} \in S$ if action $a \in A$ is taken in state $s \in S$, and $R(s, a, s^{'})$ is the reward of transition $(s, a, s^{'})$. At each time step $t$, the process is in some state $s \in S$, and an agent needs to choose a legitimate action $a \in A$. The process then moves to a new state $s^{'} \in S$ at time $t+1$ probabilistically and the agent receives a reward correspondingly. The probability that the process moves into a new state $s^{'} \in S$ is determined by both the current state $s \in S$ and chosen action $a \in A$, formally described by state transition probability $T(s, a, s^{'})$. Given $s \in S$ and $a \in A$, the probability is conditionally independent of all previous states and actions, which indicates that the state transitions satisfy the Markov property. If the time spent in each state transition is regarded as an additional parameter, a semi-MDP can be modeled. The differences between an MDP and a semi-MDP are summarized in Figure \ref{MDP}.

To improve the QoS in cellular networks, a semi-MDP is used to formulate the problem of minimizing new call and handoff call blocking probabilities in \cite{966366}. As a result, a channel allocation scheme that prioritizes handoff requests has been proposed. For QoS provisioning in wireless ad hoc networks, joint bandwidth allocation and buffer management has been considered in \cite{1424864}, where a semi-MDP is also used to model the system. Accordingly, an algorithm has been proposed to maximize the long-term reward and minimize the QoS violations. In \cite{6504467}, power control for wireless energy harvesting sensor networks has been studied. The power control for packet transmission attempts is modeled as a \emph{partially observable Markov decision process} (POMDP), which outperforms conventional models.

A simplified version of the MDP is the \emph{multi-armed bandit} (MAB), with only one state for the environment. In this case, the agent only needs to determine the best action, i.e., pull the arm. MAB has been used in \cite{4657097} to strike a balance between the exploitation of the environment and the exploration of the accumulated knowledge for opportunistic spectrum access. MAB can be further extended to \emph{multi-player MAB} (MP-MAB), where the reward of a player depends on the decisions of other players. In \cite{6939716}, a distributed channel selection problem in D2D networks has been modeled as a MP-MAB game with side information and a distributed algorithm has been proposed.

Beyond traditional MDP, RL does not require prior knowledge of the state transition probability $T$ or the reward $R$, which makes it suitable for many real-world applications. For instance, in \cite{5452965}, RL has been used for opportunistic spectrum sharing, which achieves good performance without prior knowledge on the environment. To improve the spectral efficiency in heterogeneous networks, a distributed strategy has been proposed in \cite{6994301} based on RL to reduce both intra-cell and inter-cell interference and improve the throughput under the environment uncertainty. In \cite{4428658}, distributed independent RL based on Q-learning has been used so that only local information at nodes is required and the utility value given a specific task can be optimized. To reduce the co-channel interference to macrocells, channel selection and power allocation based on Q-learning are proposed for self-organizing femtocells in \cite{bennis2010q}. In a heterogeneous network powered by hybrid energy, a model-free RL approach has been proposed in \cite{8100645} to learn the optimal policy for user scheduling and resource allocation so that the overall energy efficiency can be maximized.

\subsubsection{Feedback in reinforcement learning}
DRL is promising for intelligent perception and action due to its powerful capability of representing features. As aforementioned, the  intelligent agent gradually maximizes the long-term reward by continuously interacting with the environment. To achieve effective interactions, the agent needs feedback from the environment. In particular, the feedback can be used to evaluate the goodness of the selected action and  adjust the actions for the next step. The dominant feedback is different in single-agent scenarios and multi-agent scenarios.

\subsubsection*{Feedback in single-agent scenarios}

To begin with, we call the communication node equipped with an agent as an intelligent node. Then, the dominant feedback in a single-agent scenario is usually from conventional communication nodes to the intelligent node. Typically, these feedback is used to construct radio environment state as well as experience. By continuously using this kind of information to train the RL model (e.g., Q-table or NN), the trained model can learn whether an action is good or not at a certain moment and gradually converge to the best action policy. For example, \textit{Yu et al.} \cite{yu2019deep} have considered  a scenario, in which multiple conventional communication nodes operating different MAC protocols try to access an access point by a common channel, and an intelligent node wants to coexist with these communication nodes by intelligently making action decisions on whether to access the access point or not based on the radio environment state, which is defined as the previous action-observation pairs. In particular, the agent can obtain the previous actions since they are stored locally at the intelligent node. The observation is defined as the impact of an action on the transmission of all communication nodes. Note that this kind of information is only available at the access point. Then, the access point needs to feed such information back to the intelligent node for the action policy optimization. Besides, \textit{Yang et al.} \cite{YC_UAV_ICCS} have studied a UAV network, in which multiple UAVs act as BSs (namely, UAV-BS) to serve the ground users and each user independently selects one UAV-BS to access. One intelligent user makes an action decision on which UAV-BS to access based on the radio environment state. Note that the throughput of the intelligent node is related to the access policies of conventional communication nodes, which are unknown to the intelligent node. Then, the UAV-BS accessed by the intelligent node needs to feed the access information of conventional communication nodes in the previous time slot to the intelligent node, such that the intelligent node can learn their access policies and  optimize its  access policy.

\subsubsection*{Feedback in multi-agent scenarios}

When multiple communication nodes want to intelligently make action decisions, one straightforward method  is adopting a centralized RL agent, which is responsible for controlling all the actions of these communication nodes. In particular, the agent can collect all the related radio environment states of these communication nodes to train the RL model and make action decisions for all nodes simultaneously. Nevertheless, such a centralized scheme faces two main challenges. First, it is difficult for the agent to collect all the related radio environment states of each node through feedback in practical situations. Second, the size of the state-action space at the centralized agent will  increase exponentially as the number of communication nodes grows. A large state-action space may slower the convergence rate of the RL model since the agent needs to explore the whole state-action space for model training.

To address the above issue, a multi-agent framework is usually adopted. In the multi-agent framework, each communication node is equipped with an agent, i.e., intelligent node, and they make action decisions independently. In particular, the state-action space of each RL model is only determined by the state space and action space at each individual intelligent node and remains constant even when the number of intelligent nodes grows. For instance, \textit{Li et al.} \cite{li2009multi} have adopted the multi-agent Q-learning algorithm to solve the channel selection problem in CR systems. \textit{Bennis et al.} \cite{bennis2010q} have proposed a multi-agent Q-learning algorithm to avoid the interference in self-organized femtocell networks. The intelligent nodes in \cite{li2009multi} and \cite{bennis2010q} only exploit the feedback from the local environment (similar to those in single-agent scenarios) to make action decisions and to update Q-tables, by ignoring the action policies of other intelligent nodes. Since the reward of an intelligent node is also affected by the action policies of other intelligent nodes, such schemes may converge to a local optimum (if possible). Later, \textit{Chen et al.} \cite{chen2017echo} have suggested that each intelligent node is informed of the selected actions of other intelligent nodes through feedback in each time slot. It has been demonstrated that such design can accelerate the convergence of the RL model.

Recently, \textit{Guo et al.} \cite{Dongning_Guo_DRL} have proposed a novel multi-agent framework, which includes a centralized agent and multiple intelligent nodes. In particular, the centralized agent is responsible for training a common RL model for all the intelligent nodes and each intelligent node makes action decisions independently according to the trained RL model. In this framework, each intelligent node needs to feed the local experiences back to the centralized agent, which randomly samples the experiences to train the RL model. To make an action decision at intelligent nodes, each intelligent node needs the feedback from the centralized agent to obtain the latest RL model, as well as the feedback from the local environment to construct the radio environment state as the input of the RL model. Compared with the distributed framework in \cite{li2009multi,chen2017echo}, the RL model in \cite{Dongning_Guo_DRL} can converge at a faster speed with the cost of the overheads caused by the feedback between intelligent nodes and the centralized agent.

\begin{figure}[!t]
    \centering
    \includegraphics[scale=0.48]{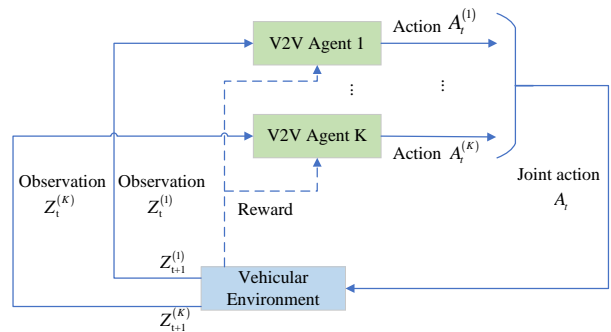}
    \caption{The multi-agent RL formulation of spectrum sharing in vehicular networks.}
    \label{V2X}
\end{figure}

Moreover, \textit{Liang et al.}~\cite{liang2019spectrum} have  developed a distributed spectrum and power allocation algorithm that simultaneously improves performance of both vehicle-to-vehicle (V2V) and vehicle-to-infrastructure (V2I) links by adopting the multi-agent model to determine the V2V spectrum sub-band selection and power control as shown in Fig.~\ref{V2X}. Particularly, the RL framework is adopted and each vehicle is regarded as an agent. For each agent, the observation is based on the environment state including bandwidth, transmission period, interference, and channel capacity, which could be updated in each transmission period. The reward is based on the achieved transmission capacity. The proposed multi-agent RL based method includes a centralized training stage and a distributed implementation stage. The overhead occurred in the aforementioned work could be reduced as each V2V agent receives only local observations of the environment  at the implementation stage. More examples of applying machine learning algorithms in intelligent vehicular networks can be found in~\cite{8633948,8472113,8345672}.

\section{Challenges in Intelligent Communications} \label{IC}
During the past two decades, we have witnessed the rapid developments of intelligent communications on CR. 
It had been listed as one of the potential techniques to be adopted at the beginning of research  for each generation of cellular networks. Even for the future 6G, \textit{Federal Communications Commission} (FCC) has recently proposed a new framework to exploit the block chain technique to manage wireless spectrum, which makes it quite promising for CR.

However, even though CR has been adopted in many standards and its standardization is also currently performed at all levels, including the ITU, IEEE, ETSI, and ECMA, we have to admit that there is still a long way for the implementation of CR.  If we are still opportunistic for CR in the future design of intelligent communication systems, we need to address the following technical challenges:
\begin{itemize}
\item \textbf{Commercialization around the world:} So far, large-scale commercial tests for \textit{TV White Space} (TVWS) have been carried out in many countries, which allows the user to temporally access TV channels that were assigned to  analogue TV signals. Moreover, UK has made the TVWS open to the public for  commercial utilization. Spectrum resources are  normally managed by national governments. Apart from the technical barriers, the governance of spectrum becomes the key to enable of the international implementation of CR. In order to further promote the large-scale implementation of CR over the world, the compatibility with existing standards and architecture should be carefully considered.
\item \textbf{Pricing and payment:} Once the green light is given to CR, efficient spectrum resources management becomes the key enabler. The aforementioned block chain based spectrum management provides an efficient and distributed approach. Moreover, as the secondary channel access period could be short and the transmitted data from secondary users could be small. The transactions could happen quite frequently. How to charge the small and frequent payment will become very challenging. The small contract in block chain might be utilized for billing the spectrum utilization cost for secondary users in an efficient and secure way.
\end{itemize}
While for the broader concept of intelligent communications, we could be  much more opportunistic. Apart from CR, the spirit of intelligent communications has spread over the design of the whole communication systems, from blind equalization to adaptive coding and modulation, over the past decades. Thanks to the great exploration of AI in recent years, intelligent communications become promising in the design of beyond 5G networks. It can be noted that perception and action are highly coupled in many of the intelligent communications. Compared to the typical methods, intelligent communications face the following challenges:
\begin{itemize}
\item \textbf{Open dataset}: Different from conventional communication technologies, intelligent communications exploit the patterns and features hidden in massive historical data for system performance enhancement. It is clear that obtaining sufficient and valid data is the precondition to realize the intelligent communications.  In many cases, data are generated by  some theoretical models through computer simulators, which may be oversimplified or inaccurate and cannot guarantee the validity of the generated data. Therefore, an open-access dataset for real and typical communication scenarios is needed for valid performance evaluation and fair performance comparison.
\item \textbf{Tailored learning framework for communication systems}: Different machine learning and deep learning algorithms have been developed for intelligent communication systems. However, most of them are inherited from the designs suiting well for problems difficult to model, such as computer vision. The existing designs mainly use the learning tools as ``black box" and may not work perfectly for the communication systems. Therefore, it is desired to develop new learning frameworks tailored for communication systems, to solve the problems in a more efficient way.
\item \textbf{Intelligence versus reliability}: We have to note that the tradeoff between intelligence and reliability of the systems has to be carefully considered. In particular, some abnormal events might mislead the intelligent systems, which could further guide the whole system into a status that makes wrong or even unacceptable decisions. Therefore, the AI-enabled intelligent communication systems should be smart and robust enough to anomaly.
\end{itemize}

\section{Conclusions}\label{Conclusions}
This article has tried to provide a brief overview on the developments on intelligent radios over the past two decades. By treating the intelligent radio as a perception-action cycle, we started from the review of typical spectrum sensing and sharing approaches. Afterwards, the recent advancements on AI-enabled intelligent communications have been characterized from the perspectives of perception and action, respectively. Moreover, it is noted that the two aspects are normally coupled, especially in the AI-enabled approaches. After reviewing the evolution of intelligent communications from cognition to AI in the past 20 years, we have discussed the potential future of intelligent communications. By identifying the barriers that block the large-scale implementation of CR, we have further discussed the challenges faced by AI-enabled communication systems. We believe the intelligent communications will be applied in many practical systems in the near future even though there are still many challenges to be addressed.

In order to make this article clean and tidy with particular focus on the milestone work in the past 20 years, we have omitted some topics in intelligent communications from cognition to AI, such as policy and standardization, spectrum usage measurements and statics modelling, and security and privacy. The readers are suggested to refer to other existing articles~\cite{Filin:2011,Bkassiny:2013,Stuber:2009,Attar:2012}.

\bibliographystyle{IEEEtran}
\bibliography{IEEEabrv,CR_AI_bib}
%

\end{document}